IAC-22-B3.IPB.2

# Electroencephalography (EEG), electromyography (EMG) and eye- tracking for astronaut training and space exploration

**Leonie Becker\*, Tommy Nilsson, Aidan Cowley**

*European Space Agency (ESA) – European Astronaut Centre (EAC), Linder Hoehe 51147 Cologne, Germany*
\* Corresponding Author: leoniebecker1010@gmail.com

**Abstract**

The ongoing push to send humans back to the Moon and to Mars is giving rise to a wide range of novel technical solutions in support of prospective astronaut expeditions. Crew capsules for long-duration travel, prospective lunar surface habitats, and extravehicular activity (EVA) suits are a few of the advanced systems set to underpin future human space explorations. Such systems are in need of efficient human-machine control mechanisms and interaction interfaces that would ensure optimal work performance while safeguarding astronauts' physical and mental well-being. Against this backdrop, the European Space Agency (ESA) has recently launched an investigation into unobtrusive interface technologies as a potential answer to such challenges. In theory, the application of these technologies could enable future crews to initiate actions without requiring any form of manual manipulation, resulting in a more intuitive, safe, and flexible control mechanism in numerous situations. Potential examples range from the control of robotic arms and rovers to the interaction with virtual and augmented reality applications. Three particular technologies have shown promise in this regard: EEG - based brain-computer interfaces (BCI) provide a non-invasive method of utilizing recorded electrical activity of a user's brain, electromyography (EMG) enables monitoring of electrical signals generated by the user's muscle contractions, and finally, eye tracking enables, for instance, the tracking of user's gaze direction via camera recordings to convey commands. Beyond simply improving the usability of prospective technical solutions, our findings indicate that EMG, EEG, and eye-tracking could also serve to monitor and assess a variety of cognitive states, including attention, cognitive load, and mental fatigue of the user, while EMG could furthermore also be utilized to monitor the physical state of the astronaut. Such capabilities may well turn out to be important enablers for the success of future space expeditions. Indeed, monitoring cognitive states could, e.g. help guide adaptive automation, improve crew mental health and concentration via neurofeedback, indicate training success, or help astronauts to optimize the timing of their actions during future missions. In this paper, we elaborate on the key strengths and challenges of these three enabling technologies, and in light of ESA's latest findings, we reflect on their applicability in the context of human space flight. Furthermore, a timeline of technological readiness is provided. In so doing, this paper feeds into the growing discourse on emerging technology and its role in paving the way for a human return to the Moon and expeditions beyond the Earth's orbit.

**Acronyms/Abbreviations**
AR: Augmented Reality
BCI: Brain-Computer Interface
EEG: Electroencephalography
EMG: Surface Electromyography
ESA: European Space Agency
EVA: Extravehicular Activity
ISS: International Space Station
MI: Motor Imagery
SA: Situational Awareness
SSVEP: Steady State Visually Evoked Potential
VR: Virtual Reality

**1. Introduction**
A revived interest in sending humans to the surface of the Moon has emerged, as the National Aeronautics and Space Administration (NASA) plans to return humans to the lunar south pole as part of the Artemis program. In the subsequent years, this will conceivably lead to the establishment of a permanent human presence on the lunar surface and eventually humanity's expansion to Mars and further into deep space. In combination with commercial space flights set to become widely available in the near future, the demand for novel technology capable of improving safety, efficiency, and reliability of future missions is currently undergoing a rapid surge. Indeed, our capacity to complete future human space flight missions in a secure, reliable, and effective manner will be essential to the overall success of future human space flight.

Even though the technological development of systems such as the eXploration Extravehicular Mobility Unit (xEMU) suit, and crew capsules for long-duration space flight has advanced significantly in recent years, future exploration missions pose unique challenges in terms of astronaut safety, performance and health. Future human spaceflight missions will expose astronauts to radiation, microgravity, and other space-related hazards to a greater extent and for a longer duration than during past endeavors in human spaceflight. Moreover, the weightless environment or the change in gravity on the planetary surfaces of Mars and the Moon require different ways of operating technologies, as the motion of the astronaut is restricted by the cumbersome astronaut suit, and motor performance is limited due to the lack of gravity [1]. Apart from dealing with considerable physical strain, future astronauts will likewise need to overcome numerous psychological hurdles, including issues with sleep [2], feelings of isolation or confinement, and interpersonal difficulties [3]. Moreover, numerous effects on mood and well-being have been documented, such as an increase in anxiety, anger, fatigue, and depressive symptoms [4]. Notably, research demonstrated that astronauts' decision-making abilities and alertness can be impaired by their weightlessness condition and by excessive workload [5]. In situations of excessive workload and decreased alertness, limited resources and abilities contribute to a significant decline in performance which, in turn, might lead to the failure to complete assigned tasks [6]. In contrast, particularly during long-duration space flights, low levels of workload due to high levels of automation, resulting e.g. in boredom, may arise [4]. Whereas excessive workload, as well as extremely low workload levels, have been linked to a phenomenon known as *a loss of situational awareness (SA)*, in which the astronaut is unable to react appropriately in a high-risk situation where his input is required [7]. Notably, during long-duration space flights, arising challenges, such as unexpected critical situations, have to be resolved mostly by the astronauts themselves, as a return to Earth in the event of an emergency is not possible and, depending on the location of the astronauts, a communication delay between mission control and the astronauts has to be expected [3]. In addition, missions to the lunar surface or Mars can last months or even years, which requires a high level of preparation and training. Therefore, astronauts need to be able to evaluate their own health status and performance [8].

Here, especially the use of new technologies could prove beneficial to meet those needs, i.e. to ensure good crew performance in stressful and new environments and to inform the crew about possible health and performance issues that may arise during long-duration missions [8]. In light of this, our paper should investigate the question of how unobtrusive interface technologies, specifically the utilization of electroencephalography (EEG), electromyography (EMG), and eye-tracking technology could be a potential solution to these issues.

EEG, EMG, and eye-tracking technologies could not only help to mitigate some of the issues associated with future human-space flight, but could also give further insights into the neurocognitive and neuropsychological parameters that are affected by space flight [9], but their application could theoretically enable future crews to initiate actions with less or no manual manipulation, resulting in a more intuitive, secure, and flexible control mechanism in conditions of weightlessness, high workload, and limited flexibility. In addition, psychological challenges could likewise be mitigated using eye-tracking and EEG or EMG recordings, as such technologies permit the monitoring of a variety of cognitive states, including attention, cognitive load, and fatigue, all of which are known to be negatively impacted during space missions. Here, it is possible to determine the need for a break, the effectiveness of training, or the optimal mission time-point.

In the next sections, a brief definition of each of the three proposed technologies is provided, followed by a detailed explanation of possible applications in the context of future human space flight. In that respect, the paper is thought of as an extension and update to the findings of [10]. The paper closes with a discussion and an overview of the technological readiness level of each proposed technology.

## 2. BCIs

BCI technology utilizes recorded brain activity to transmit commands and/or messages to a computer system, which then analyzes the current brain activity using feature extraction and classification techniques. In a subsequent process, output signals or feedback are generated that can be used for a variety of applications, such as the estimation of an astronaut's cognitive state or the control of technical systems [11].

Therefore, the recorded brain activity can be analyzed and further utilized without requiring any movement by the astronaut. Notably, a distinction can be made between *active, reactive,* and *passive* BCI technologies. Active BCIs require mental effort and intentional thoughts by the user to operate an external system. Reactive BCI technology relies on the perception of external stimuli that elicit a specific reaction in the electrical activity of the brain, such as visual signals. Lastly, passive BCIs utilize passively recorded brain activity of the user, such as indicators of fatigue or workload [11].

### 2.1. Non-Invasive EEG Recordings for BCIs

EEG uses *electric-magnetic responses* to estimate brain activity, whereas the measurement takes place by recording electrical signals of a large number of neuron clusters in the brain. Non-invasive EEG allows for the analysis of electrical activity recorded from the scalp of the participants. Even though only a lower spatial resolution ( that is the detailedness of the measurement) can be achieved, EEG electrodes can be easily attached to the subject's head and do usually not require extensive training or costly equipment [11]. In spite of the lower spatial resolution, non-invasive EEG is arguably the most popular enabling technology for BCIs due to its safe and simple handling, relatively quick setup, and high temporal resolution.

For this reason, EEG recordings were analyzed aboard the International Space Station (ISS) to investigate the effects of the space environment on the human brain and its functioning, such as the influence of long-duration weightlessness on the 3D perception of spatial cues [13] and sleep [5], among others. Furthermore, in a recent private mission to the ISS, the company "brain.space" tested a specially developed EGG headset to investigate the effects of microgravity. The device does not require any conductive gel or elaborate placement of single electrodes, which may drastically increase the usability of the technology [14]. Given that EEG data promises substantial benefits, the following section proposes potential general applications for future human space flight missions.

*2.2. Monitoring of Cognitive States Using Passive BCIs*

*2.2.1 Mood*

Importantly, one notable application of EEG recordings is their utilization to assess the astronaut's cognitive or psychological state during future human space flight missions. During future human space flights, astronauts likely face feelings of isolation or confinement, as well as interpersonal difficulties that may lead to negative psychological outcomes, such as a low mood, depressive symptoms, and elevated levels of stress or anxiety [3], [4]. Here, research has demonstrated that EEG data could be successfully used to detect signs of each of the aforementioned cognitive states. For instance, [15] used EEG data to identify signs of depression, whereas a review by [16] demonstrated that EEG can be used to identify various emotional states, such as anger, happiness, fear, and relaxation. In addition, stress levels [17] and anxiety [18] were detected in prior research. Once a negative psychological state is identified, the astronaut may seek various options, such as taking a break, engaging in therapy programs, or in relaxation techniques.

Likewise, a specific EEG technology, known as neurofeedback, could be utilized as a further treatment option. Neurofeedback applications provide users with real-time feedback on their cognitive state, enabling the user to self-regulate brain activity to achieve the desired state. Thus, the user can gradually develop a strategy to allow a modification of one's own brain- activity based on a trial and error approach [11]. Neurofeedback applications have already been applied successfully to treat a variety of psychological disorders and adverse psychological states, including depression, anxiety [19], and stress [20]. A key advantage of neurofeedback applications is the fact that they can be employed without requiring any external help from ground control, as they can be successfully utilized by the astronauts themselves, which could be especially relevant during long-duration missions to Mars or Moon.

*2.2.2 Sleep, Fatigue, and Alertness*

Importantly, human space flight has also been associated with negative effects on sleep. Research in this domain has demonstrated that sleep deprivation and a change in the circadian rhythm can be anticipated, followed by fatigue and drowsiness [21], [22]. To illustrate, [2] analyzed the sleep patterns of astronauts during different space shuttle missions and aboard the ISS, revealing a significant reduction in sleep time despite the astronauts' frequent use of sleep-promoting drugs. Importantly, prior research could demonstrate that most people are unaware of the negative effects of sleep deprivation on cognitive performance measures, highlighting the need for objective measurements to inform the crew about their current status [23]. To further investigate the effects of space conditions on astronauts' sleep, a team of researchers from the University of Aarhus is currently developing an ear-worn EEG device (ear- EEG), which is designed to measure sleep patterns aboard the ISS. As the system can be worn similarly to an earphone, there is no need for a large EEG- cap, thereby minimizing any potential impact on sleep quality caused by the device [24]. In future space flight missions, such a system could not only be used to study sleep patterns in space, but also to inform astronauts of less-than-ideal sleep patterns so that they could take corrective actions. Additionally, research could demonstrate the efficacy of neurofeedback applications that can be used to improve total sleep duration [25], possibly mitigating the adverse consequences on sleep due to the environment in space.

Moreover, space conditions have been linked to an increase in fatigue-related symptoms [4], while a decrease in alertness can be observed [5], which have been both linked to decreased task performance. Due to the importance of avoiding the negative effects of fatigue and decreased alertness in a variety of domains significant progress has been made in this area. For example, [26] utilized EEG data to correctly estimate alertness and drowsiness levels with an accuracy of above 95%. Other research could demonstrate that EEG signals can not only be used to predict driver fatigue with a high degree of accuracy but also future task performance [27]. In the context of future human spaceflight, drowsiness, alertness, and fatigue could be used to alert astronauts of their condition. Therefore, mission- breaks can be implemented or missions can be rescheduled based on the current and predicted status of the astronaut.

*2.2.3 Workload, SA, Boredom, and Attention*

Another important factor during future human space flight missions is that the astronauts will likely face both extended periods of boredom, such as during heavily automated phases of long-distance flights, as well as instances of extreme high workload and stress, such as during EVAs or teleoperation that require the full attention of the astronaut. Using EEG – based BCIs could allow for a risk reduction in these situations through the surveillance of the aforementioned states. This aspect is especially relevant as unhealthy workload levels (e.g. boredom or excess workload) run the risk of causing a loss of SA and reduced performance and safety during critical phases of space flight missions. As such, they should be prevented at all costs. In this regard, a review by [28] demonstrated that EEG has been successfully used to estimate workload levels for a variety of applications. Moreover, e.g. [29] utilized EEG data to assess attention levels in a driving task. In future space flight missions the recorded data can be utilized in a variety of ways: 1) Additional tasks, such as training sessions or additional leisure activities could be introduced in cases of boredom to alleviate low workload situations. 2) The astronaut may take a break, another astronaut may take over or assist in completing the task, or the mission may be rescheduled if the workload is excessively heavy, attention levels decline, or SA is compromised. 3) Neurofeedback training could be used to enhance cognitive and sensorimotor functioning and to improve working memory, as well as attention [30]. 4) Adaptive automation, which bases the level of autonomous support provided by a technological system on the cognitive state of the astronaut, could be implemented while preventing mental over-, as well as under-load and ensuring high SA and performance benefits [32]. [33] demonstrated that the degree of support in an air traffic control task can be modulated using an EEG- based BCI, resulting in optimal performance. Adaptive automation solutions would therefore be an ideal tool for human spaceflight missions, as they could be used for tasks that normally require a high degree of autonomous control, such as the monitoring of various systems, docking procedures, and telerobotic operations. Moreover, adaptive automation could also help to prevent astronauts from losing crucial skills that could lead to a decline in performance [7].

*2.2.4 Effectiveness of Training and Design Evaluation*

Importantly, research demonstrates that especially continuous training is crucial for high performance. The estimation of objective workload levels and related concepts such as SA or attention might also be helpful during training and development scenarios. In that respect, [33] described how " performance assessment metrics used during training do not explicitly account for their cognitive workload while performing a task (…) [, which] leads to an incorrect assessment of operators' abilities" (p. 51). To counteract this effect, [34] used EEG measurements during a simulated on-orbit telerobotic training for the Canadaarm2 attached to the ISS, during which time-pressure and task difficulty were modeled to alter workload levels. In a subsequent step EEG - based BCI technology could be used to ensure that astronaut training is always adjusted to the workload level of the astronaut, while preventing overly difficult scenarios and easy tasks, by triggering different scenario parts or events, which may prevent excessive cost and will likely lead to a reduced training time [35]. In the context of astronaut training, not only the surveillance of workload estimates can be helpful, but data about attention and concentration levels could also be used to predict the optimal time-point or schedule for training.

In training scenarios also the utilization of neurofeedback applications could give the astronauts an estimation of their own workload and attention levels, which could aid them in the perception of their own abilities. Importantly, neurofeedback applications could be used in the context of virtual reality (VR) settings, which can reduce training costs significantly. Moreover, several objective indicators, such as workload or SA can be used to evaluate designs of technological solutions, such as interfaces or procedures that need to be completed during space missions, without distracting the user [36].

*2.3 Control Input*

Undoubtedly, future human spaceflight missions will expose astronauts not only to psychological and cognitive challenges, but they will also encounter physiological hurdles through high levels of radiation, microgravity, and other space-related hazards that could hamper the performance and safety of astronauts. Particularly the weightless environment or the difference in gravity requires new ways of operating technologies since the astronaut's motion is impaired by the bulky astronaut suit and motor performance is impeded owing to lack of gravity. There are numerous instances in space where moving the hands for controlling input modalities, such as when using touch displays or a joystick, may not be easily possible (e.g. EVAs). The selection of options without requiring active sensorimotor performance might be more intuitive and straightforward and may therefore be associated with greater usability and less mental effort, as direct communication without any output delays is possible, while BCI technology allows for a translation of actions based on intentions that do not need to be translated into (complex) motor actions in the first place. This aspect may be especially relevant when astronauts are exposed to harsh conditions in space, where direct mental teleoperation of external robotic agents could be quite advantageous, especially as manual control is generally associated with a delay in output in microgravity conditions. Therefore, the use of BCIs as a way of interacting with technologies could be a potentially faster and easier way of conveying commands to various technologies. In this regard, it may be possible to not only augment the capabilities of the astronaut by more intuitively operating a rover or robotic arm by brain commands but also to achieve higher degrees of multitasking [37].

*2.3.1 Reactive Steady State Visually Evoked Potential (SSVEP) - Based BCIs*

When a user observes a light stimulus that flickers at a constant frequency (between 3 and 40 Hz), the electrical activity of the visual cortex synchronizes with the stimulus's frequency and its harmonic frequencies. The stimulus elicits a so-called SSVEP that can be measured via EEG electrodes. Various visual stimuli, indicating options or actions, can be represented by flashing patterns with varying flicker frequencies. By concentrating visually on one of the flickering stimuli, the user can select a particular option. SSVEP recordings, in contrast to eye-tracking, only require a shift in attention or "visual interest", therefore a shift in gaze is not compensatory to select options. In general, the flickering stimuli needed to elicit an SSVEP response in the brain can be represented through basic shapes, such as geometric forms (circles/squares) that can be displayed on displays or screens [11]. Studies investigating the use of SSVEP -based BCIs could achieve quite high classification accuracies of around 95% for 25 different stimuli [38]. Importantly [39] demonstrated that SSVEP- based BCIs successfully worked in space conditions aboard the Tiangong-2 Space Lab, where two astronauts achieved comparable performance as on ground while achieving similar accuracy levels as [38] for 4 different stimuli.

Even though SSVEP-based BCIs are associated with a number of advantages, such as a short training period due to a small inter-subject variability, ease of use and the demonstration of their reliability in weightless conditions, their real-world application in the context of space is questionable. Current state-of-the-art systems require relatively large stimuli to elicit strong SSVEPs in the visual cortex of the subject in order to improve classification accuracy, and the number of possible stimuli is constrained by the number of frequencies that can be used to elicit useful responses in the visual cortex [40]. Moreover, the use of SSEVP- based systems require the visual attention of the astronaut. Hence, options that do not lie within the visual range of the astronaut cannot be selected. Furthermore, it is questionable, if negative consequences on important cognitive factors such as attention or workload through the additional flickering visual stimuli can occur, which also have been frequently reported as being perceived as uncomfortable after some time. The relatively long classification time of around 1.7 seconds that was, for instance, achieved by [38] could impact usability and security for the astronauts, also considering that currently used technologies, such as touch display interfaces take significantly less time to react [41]. Moreover, a vulnerability to artifacts created through speaking (counting loudly) and thinking (mentally counting) could be shown [42]. Other researchers used shared control approaches to improve the accuracy of the system that allows for the co-joint exertion of commands by both the operator and an automated system can enhance system accuracies. Whereas shared control systems could overcome "individual shortcomings of pure teleoperation, or autonomy." (p. 1310) [43]. One might think of an application where the technical system corrects the operators' commands or vice versa.

Even though currently significant challenges are associated with the use of SSVEP- based BCIs, several different possible applications are thinkable in the distant future. During future EVAs or during lift- off an AR system could be integrated into the astronaut helmet's visor, allowing for the selection of various options. The astronaut could communicate operation signals to a rover, activate communication, or turn on/ off lights, navigational cues or emergency functions.

Nonetheless, because SSVEP-based BCIs rely on the input of external visual stimuli, there may be a risk of selecting the incorrect or no option by visually focusing on a nearby region. Therefore, it remains questionable whether SSVEP-based BCIs can be used without an additional safety function, such as a confirmation button press. In addition, buttons could be used to activate the appearance of the stimuli, reducing the duration of exposure to flickering stimuli. Even though this would still require the presence of another input method.

In terms of astronaut training, [44] could demonstrate that SSVEP- based BCIs could also be implemented into virtual and augmented reality (AR) applications for training scenarios. In this context, SSVEP- based BCIs could be a more intuitive way to move through a virtual environment or to activate actions in VR or AR. Such systems could also be useful during space missions for additional training, to display instruction, or for leisure activities. This use case may be more realistic in nearer future, especially as there are fewer safety-relevant aspects involved in training scenarios. In this regard, the use of SSVEP- based BCIs also gained attention in the commercial domain, where the company *NextMind* developed a BCI system to control visual interfaces e.g. in the VR and AR realm [45].

*2.3.2 Active Motor Imagery- based (MI) - BCIs*

MI BCIs select actions based on the recorded activity of the sensorimotor cortex. The technology makes use of the fact that brain activity, elicited when performing an actual movement, can also be elicited by simply thinking about or mentally rehearsing a particular movement. Consequently, actions such as picking a specific target or commanding a robot can be associated with imagined body movements (e.g. moving one arm to the right). One of the greatest advantages of this technology is that it does not rely on external stimuli, such as visual patterns, which could be unpleasant or prone to technical failure (e.g. display error). In addition, the participant has complete control over the imagined body movements that can be paired with a specific action. Furthermore, allowing the user of the BCI system to make active decisions at their own pace regarding when to activate or select a particular option promises a more intuitive and natural method of interacting with a technological system [46].

Nonetheless, in general, the approach necessitates more complex machine learning algorithms and longer training phases due to the increased inter-subject variation of the signal. Yet, in recent years promising results could be achieved. A recent review by [47] demonstrated that overall high accuracy levels of around 83%–93% for applications with multiple mental commands can be currently attained. Even though these studies show initial promising results, [48] demonstrated that the classification accuracy of MI BCIs significantly decreases in scenarios where external "noise" (e.g. eye movements, audio signals, speaking) is created that would likely be present in real-world applications. Yet, importantly, two studies demonstrated that the use of a BCI based on MI feasible in microgravity conditions. First, [49] instructed participants to move a balloon to the left or right on a computer screen during a parabolic flight. Second, [39] demonstrated the feasibility of MI BCIs in space conditions aboard the Tiangong-2 Space Lab, while achieving accuracies of 74% to 87% during the classification of two different stimuli (left/right arm movement).

Feasible applications during future and longer space missions would be the intuitive operation of a robotic arm or spacecraft using motor- imagery tasks, whereas the operation is not reliant on a physical location to issue control commands. Therefore, sending an emergency message, and activating options such as the autopilot is feasible without being at a specific required space. Yet, even though the aforementioned experiments in space have demonstrated that MI BCIs can be used in microgravity conditions, it remains questionable whether any error rate can be tolerated in a domain as safety-critical as human space flight. Moreover, long training times [50] raise the question of whether such a system would currently reduce astronauts' mental workload and improve usability in comparison to conventional input modalities, such as touch displays or keyboards that generally do not require any training. Additionally, the prolonged use of an EEG cap can be uncomfortable. Not only technical aspects need to be considered in that respect: a number of physiological issues could be associated with the processing of astronauts' electrical brain signals that may necessitate a change or update in training data used to develop BCI technology on earth [37]. Importantly, for more complex tasks, such as operating a robotic arm, a high transfer rate of commands would be necessary to successfully complete tasks that can currently not be reached [51]. Although in this regard, shared control approaches could be able to compensate for this aspect at least partly [52].

**3. Eye Tracking**
Eye-tracking technology can be used to assess a variety of cognitive states and to transmit commands to a technical system by monitoring the position and movement of the user's eyes. Aboard the ISS, eye-tracking measures have been used to investigate the effects of microgravity on the vestibulo- motor system, as astronauts frequently experience space motion sickness and require an initial period of adaptation to the microgravity environment upon arrival at the space station [13]. Eye-tracking technology has a long history of use in psychological and medical research, as well as interactive applications that use the user's gaze as a control mechanism [53], which could also be applicable to a number of human space flight-related activities. In contrast to EEG-based technologies, eye tracking is considered to be more comfortable due to the fact that various implementations, ranging from wearable (e.g. eye- tracking glasses) to remote eye-tracking systems (e.g. integrated into interfaces) [54], are either easy to wear and lightweight (wearable trackers) or do not touch the user at all (mobile trackers). Although some measurements (e.g. pupil dilation) have shown to be susceptible to the influence of lighting conditions. In the realm of space flight, it is conceivable that eye-tracking systems could be easily and inexpensively integrated into EVA's suits or into various interfaces, such as displays or working stations, such as the control station used to operate the Canadarm2 on board the ISS, while preventing the astronaut from experiencing additional strain.

*3.1 Psychological States*

To estimate cognitive and psychological states, various eye metrics have been identified ranging from visual fixation or dwell times, to saccadic movements, the blink rate of the eyelid, or the response of the pupil, such as the pupil's dilation [55]. The measurement of the reaction of the users' eyes happens passively, therefore no active participation of the operator is required (in contrast to e.g. active BCI's), limiting physical strain, while ensuring that the main task stays uninterrupted [56].

### 3.1.1 Sleep, Fatigue, and Alertness

As research indicates that humans are generally unaware of their own level of sleep deprivation, circadian misalignment, and fatigue, all of which have been associated with a significant decrease in cognitive and sensorimotor performance [23], [57], there is a high demand for objective sleep quality measurements that can indicate levels of sleep deprivation and fatigue in future human space flight missions.

In this regard, [57] measured complex oculomotor signals after induced sleep deprivation using eye-tracking technology and successfully identified a distinct pattern indicative of such adverse states. In addition, various eye-tracking measurements are frequently used to assess fatigue and drowsiness. Importantly, a meta-analysis by [58] demonstrated that the analysis of saccade (rapid movement of the eye between fixation points) mean- and peak- velocity, in particular, promises a reliable and rapid measurement of mental fatigue, whereas both measures decrease significantly with increasing levels of fatigue. They concluded that these measurements provide a quick and accurate measurement also outside the laboratory. Importantly, through interpolating missing saccade and fixation data, [59] could show that Hierarchical-Based Data Analysis approaches promise the estimation of fatigue levels also when only low-quality eye-tracking (e.g. noise through light reflections or movements) data is available.

In general, research in fatigue and alertness detection systems based on eye-tracking is already so far advanced that it is actively implemented in a variety of cars available to the public [60]. With regards to future human space flights, such systems could be used similarly to the driving context, e.g. to alert the astronauts of their status. Therefore, they can take a break or re-schedule the mission. Moreover, eye-tracking data could give further insights into fatigue-evoking situations during long-duration space flight missions, such as EVAs.

### 3.1.2 Mood

In contrast to BCI technology, eye tracking-based systems are currently not sufficiently advanced to recognize more complex emotions correctly and only limited research has been conducted to improve the current state-of-the-art. Here, most studies classify emotions only based on their valence (pleasantness) and arousal level (emotion intensity). Especially the recognition of negative emotions has shown to be challenging, which could be especially relevant to counteract adverse states during long-duration space flights [54].

However, notably, some progress has been made to detect levels of anxiety and stress: [61] could show that self-reported anxiety levels correlated with visual fixation duration and heightened saccade rates in a simulated piloting task. [62] investigated visual dwell times and visual scanning entropy, that is the degree of randomness in visual scanning, during a simulated aircraft landing scenario. They could demonstrate that participants enhanced their visual focus on the world outside the simulated plane, which contained little task-relevant information, in cases of heightened anxiety levels. The shift in visual focus appears to be coupled with enhanced randomness in visually scanning the environment, which are indicators that attention is shifted away from goal-directed behavior to attending to other dominant or task-irrelevant stimuli. Importantly, visually attending to important elements within the environment is necessary to ensure SA as a whole and can thus be an initial indicator of it [63]. In the context of human space flight missions, eye trackers could be used to inform the astronauts about anxiety and stress levels, so that countermeasures can be taken. Additionally, information regarding the emotional response of astronauts during simulated missions could be used to assess how realistic the simulation is perceived, as simulated and real scenarios should elicit the same emotional responses [61].

### 3.1.3 Workload, SA, Boredom, Attention and Effectiveness of Training and Design Evaluation

Notably, research has shown that several characteristics that can be measured using eye-tracking technology can be utilized to measure SA and visual attention to goal-directed stimuli directly. In this regard, [64] utilized eye-gaze patterns of novice and experienced pilots to develop a system that can estimate SA. Moreover, [65] demonstrated that there is a significant link between the number and time of visual checks on speed and distance information provided that are critical to completing the task with high accuracy and performance during a manual spacecraft docking maneuver, possibly due to ensuring high SA during task completion. They proposed that visual gaze data could be used to enhance learning applications by providing feedback regarding the visual focus on various aspects of the environment or the interface. This feedback could then be used in a subsequent step to identify potential improvements, such as a more frequent and prolonged concentration on the most vital information pertinent to task completion to ensure SA. These findings are also underpinned by the results of [66], who could show a distinct visual scanning pattern of experts during a space shuttle simulation that could be used to guide training for novice users.

In the realm of detecting workload levels, [67] recently utilized eye-tracking technology to successfully analyze mental workload through recorded eye gaze fixation duration, saccade frequencies, saccade durations and recorded pupil diameters in a simulated teleoperation task of the Canadaarm2. [68] measured visual scanning entropy in a simulated aircraft piloting task that involved the completion of emergency procedures to enhance workload. They could demonstrate that visual scanning entropy decreases with increasing workload, which is generally the exact opposite of levels of anxiety and stress [62]. Furthermore, during the ESA PANGAEA – X campaign [69] pupillometry data was used to estimate workload levels during an analog mission. Notably, their findings highlight the importance of objective workload measures, as the pupillometry measures differed from the subjective workload measures indicated by the members of the team.

[70] suggested that such a system could be further utilized to flexibly determine the degree of automation (adaptive automation) to avoid too high or low task demands. During future human space missions, adaptive automated systems could be equipped with eye-tracking technology to flexibly adjust workload levels to the needs of the astronauts during piloting or teleoperation tasks. Here, several different eye-tracking measures could be implemented that can detect several different states at once (e.g. anxiety/ stress, fatigue, and high workload) and could thus flexibly adjust the level of support that is needed to complete the task successfully, as well as to inform the astronaut about adverse states, such as fatigue so that a break or external help can be scheduled.

Likewise, eye-tracking has a long tradition as a technology for usability evaluations, as analyzing the visual focus of crew members can give important insights into how interfaces should be designed. In this regard, [71] utilized eye-tracking technology to estimate gaze duration and fixation times on various interface parts during a simulated manually controlled rendezvous and docking of space vehicles. Based on their findings they suggested that especially the visibility of the velocity display should be enhanced, as it provides highly important information for the completion of the docking maneuver (indicated by high visual fixation times by the participants). Therefore, eye-tracking technology can be used to guide the design of future crew- capsules, space stations, or other relevant technologies that are needed in the context of future human–space flight and could allow for increased usability and safety. Here, especially emergency situations could be elicited that pose an objective verifiable high workload or create heightened anxiety levels.

Additionally, eye tracking technology has also been already successfully implemented in various VR head-mounted displays (e.g. HTC VIVE Pro Eye) and AR technologies such as the Microsoft HoloLens 2 that could be utilized for training, research and development purposes. In design evaluations or simulations, eye-tracking data could be analyzed to estimate the aforementioned states. The results of the analysis could then be used for future research on these constructs or to evaluate the system's design.

*3.2 Control Input*

Eye-tracking technology could also serve as a novel interface between astronauts and various technologies. Typically, our eye movements are more rapid than our actions and thus precede them. Therefore, gaze-based interaction offers a potentially more efficient and intuitive method for selecting options. To prevent accidental selections, various "confirmation" options can be implemented, ranging from the use of visual dwell time or a specific number of blinks to the implementation of a second input modality (e.g. confirmation key press, voice-based confirmation) [53].

Primarily developed for disabled users, eye-tracker-based systems have already been integrated by Microsoft in the context of a hands-free keyboard and computer mouse in their Windows 10 and 11 computer operating systems. Moreover, eye tracking-based input systems have been studied in other domains, such as piloting tasks, air traffic control, and teleoperation of robotic systems. [72] investigated gaze point and dwell time as a potential intuitive input for air traffic controllers that permits hands-free interaction with their interface. In the teleoperation domain, [73] utilized gaze and hand movement recordings to allow for an auto-corrected more intuitive operation of a robot during a teleoperation task by adjusting the angle of the camera attached to the robot, while ensuring better eye-hand coordination.

Eye tracking technology could also be an alternative "hands-free" input control technology that can be tested in a cost-effective way using AR and VR technology that can be used in the context of future human space flight training or development of applications.

**4. EMG**
EMG utilized the activity of electrical signals during voluntary or involuntary muscle contractions. In the context of space missions, sEMG, which utilizes electrodes that are placed on the skin surface to measure the electrical activity of the muscle, provides a viable solution for several applications, as it yields a non-invasive method to estimate several physiological and psychological factors. SEMG measures are, in contrast to eye-tracking or EEG-based systems, not as susceptible to changes in lighting conditions or noise, but require the correct placement of the electrodes needed to measure changes in electrical activity of the muscle [74].

### 4.1 Psychological / Physiological States

Research has begun to investigate the use of sEMG signals for the recognition of emotions. [75] classified emotions into the categories relaxed or angry based on the measurement of an inexpensive forearm sEMG electrode. Yet, most emotion recognition research seems to focus on recorded facial EMG activity (e.g. [76]). However, the usability of such a facial electrode-based system for independent use and use over longer times remains questionable.

Notably, research demonstrated that EMG electrodes attached to non-facial muscles ( e.g. trapezius or erector spinae) can effectively be used to estimate stress [77] and anxiety levels [61]. By consciously releasing muscle tension, EMG-based biofeedback has been shown to be an effective treatment for anxiety symptoms [78]. Furthermore, an increment in mental workload levels has been linked to an increase in the activity of various muscles [79]. Nonetheless, EMG measurements are most frequently used to evaluate physical demand and fatigue. In this regard, the relationship between physical and mental demand may be especially important as a high physical workload has been linked to a decrease in mental performance, and especially EVAs have been shown to be particularly physically demanding [80].

In the space flight context, [81] developed a textile wearable solution of an EMG- combined with a Near InfraRed Spectroscopy (NIRS) sensor to measure muscle activity and oxygenation for ESA. The system was implemented in a fitness tight to surveil and adjust exercise programs aboard the ISS. [82] proposed the development of the *XoSoft Gamma exosuit* for ESA, which included EMG measurements as one sub-system to alert the astronauts to take countermeasures when adverse effects due to weightlessness are detected or to artificially generate resistance to simulate gravity conditions.

Therefore, a potential area of application in the context of future human–space flight would be the integration of sEMG sensors inside the astronaut suit or their daily clothing such as proposed by the aforementioned concepts to 1) objectively record and research physical fatigue development of astronauts during IVA and EVA activities, which could be equipped with additional sensors to furthermore measure workload, stress and anxiety levels without interrupting the daily activity of the astronauts. 2) To take countermeasures, such as interrupting the given EVA activities to prevent overexertion and possibly adverse health effects or negative influences on cognitive performance during safety-relevant tasks or to implement biofeedback applications to decrease anxiety and stress levels. Or lastly, 3) to adjust and surveil exercise training plans of the astronauts to prevent physical deterioration during long-duration space flights based on the recorded data.

### 4.2 Control Input

SEMG cannot only be used to record and analyze the electric activity of various muscles but voluntary muscle activity can also be used to control various technologies. Traditionally, sEMG has been most utilized in the development of artificial limbs, but research has recently focused on alternative applications, such as telerobotics and human-computer interaction: [83] recorded muscle activity from different body parts that could be flexibly selected by the participants for the completion of a 2D aiming and tracking teleoperation task in weightless condition during parabolic flights. During the experiment, the participants did not have to move a complete body part but had to activate a specific muscle to select certain options or actions. Importantly, their findings indicate that the sEMG signal is not significantly affected by microgravity conditions and requires only little equipment.

Importantly, the promising potential of such sEMG devices was also recognized by several companies, such as Thalmic Labs Inc. which developed a cableless sMEG wristband for intuitive gesture control of robots or other technologies. Since its development, the wristband was used in several studies (e.g. [84]). Moreover, Meta recently announced the development of a wristband to facilitate intuitive interaction with AR technology. According to Meta, sEMG enables an effortless, highly reliable, and faster interaction and they intend to improve their system to the point where it can read user intentions without requiring actual hand movements [85].

Nonetheless, a review by [74] concluded that, despite the fact that motion intention prediction requires minimal user attention, electrodes that do not move or lose contact with target muscles to prevent distortion of muscle signals need to be developed. Moreover, the inter- and intra-subject variability in sEMG data are generally high, necessitating the development of systems that can be utilized in an online "real-world" scenario.

A possible promising area of application for sEMG during future human space flight missions would be the use of exoskeletons that can be e.g. used for teleoperation in space. In this regard, [86] developed an arm-based exoskeleton that could be utilized to operate humanoid-like robots during EVA's on-board the ISS and beyond, limiting possible safety and health risks for astronauts, while profiting from human performance in cases where fully automated tasks are not possible. Such a system could be further equipped with sEMG sensors to allow for less physically demanding tasks through the detection of movement intentions. This concept could be furthermore extended through the use of textile-based wearable suits and VR/AR technology that could possibly lead to higher usability and more flexibility during teleoperation scenarios, while also making the surveillance of physiological and psychological states possible (e.g. workload, stress, physical exhaustion).

## 5. Discussion and Conclusion

Future human spaceflight missions will expose astronauts to greater and longer-lasting physical (e.g., weightlessness, radiation) and psychological challenges (e.g., sleep deprivation, extremely high or low workload, fatigue, isolation) than previous human spaceflight endeavors. During future long-duration space flights e.g. communication delays and a restricted bandwidth between ground control and astronauts are likely [3]. Therefore, the monitoring of one's own health status and performance can be crucial to ensure good task performance, safety and thus mission success [8]. In this paper, the use of three enabling technologies, namely EEG recordings, eye-tracking, and EMG, has been specifically investigated. Moreover, our paper demonstrated that all three technologies can be used to initiate actions with minimal or no manual manipulation and to adjust the level of automated support (adaptive automation), for a variety of technologies, such as rovers, robotic arms, exoskeletons, or as a new form of a control interface for cabin instrumentation in conditions of weightlessness and constrained movement flexibility, during IVAs and EVAs. To estimate the technological readiness of space-related technologies NASA developed a 9 - point scale [87]. Fig. 1 provides an overview of the different technology readiness levels that can be associated with each of the three technologies and applications.

It could be demonstrated that all technologies have been shown to work reliably in weightless conditions. Moreover, the estimation of basic cognitive parameters, such as mental workload, stress, and fatigue are already in an advanced state. Given the currently greater comfortability and short training time of eye-tracking technologies, which can be worn as glasses or integrated into displays, workstations, VR and AR technologies and possibly in an astronaut's helmet during EVAs, it appears that eye-tracking currently provides a higher degree of usability for near- future human space flight missions than passive EEG- based BCI's, particularly during long-duration missions that are likely to involve an extended period of technology use. The paper also demonstrated that eye-tracking can be used to decode astronauts' visual scanning patterns, which could provide important insights into the design of future space flight technology, as well as serve as a training device for novice astronauts, improving their performance based on expert visual scanning strategies. Furthermore, eye-tracking measurements could be supplemented by wearable textiles with integrated EMG sensors, which could provide information about the astronaut's psychological as well as physical status. One of the most promising applications of workload, SA, and fatigue estimation is not only monitoring these states for research or informing astronauts to take breaks but also using these measures to guide adaptive automation schemes that adjust the amount of automated support based on the astronaut's measured workload and performance. As a result, task underloading and overloading can be avoided while maximizing performance without jeopardizing skills that may be needed in an emergency situation. Future research should thus assess whether adaptive automation technology based on eye tracking and EMG measurements, in particular, can be used reliably in weightless conditions and for longer periods of time during future missions. Additionally, other information gathered via camera recordings could be moreover utilized for additional analysis of e.g. emotions based on the analysis of facial expressions or body posture [16].

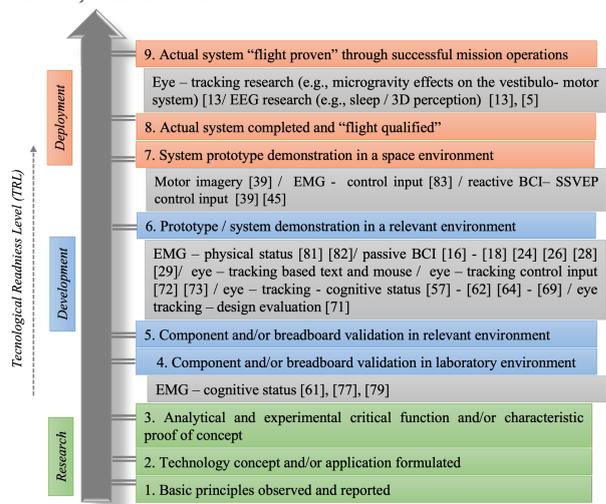

Fig. 1. TRL level of various EEG, eye-tracking, and EMG technology applications

EEG-based systems, on the other hand, promise even more advanced evaluations of psychological states in the future, ranging from the detection of various and complex emotions to workload, fatigue, and sleep-related measurements. Despite significant progress in developing commercial EEG systems utilizing dry electrodes and providing higher usability (e.g., Emotiv, Muse, etc.), longer training times and limited research in real-world settings necessitate more extensive research to achieve real-world applicability. The use of so-called ear - EEG systems, which are planned to be used to investigate sleep patterns aboard the ISS, is one promising candidate for significantly improving the usability of EEG systems during long-duration space flights: In this regard, for instance, [88] utilized ear- EEG technology to successfully estimate the attentive state of the user. However, more research is needed to enhance the applicability of ear- EEG systems to also assess other concepts, such as complex emotions or workload estimations. In the distant future, these systems could also be used to train machine learning algorithms to predict the cognitive states of astronauts, allowing for more efficient mission or training session scheduling.

One promising application of all three technologies is neurofeedback or biofeedback applications, which have proven to be an important area of research that should be extended to the realms of future human space travel. These applications could be an effective tool to counteract sleep deprivation, reduce stress, symptoms of depression, and anxiety, as well as to enhance attention [3], [19], [20], [30], [89]. Neuro- or biofeedback would yield an objective and, most notably, independent technology that astronauts themselves could use to counteract these negative states. Yet, also in this domain, more research is needed to establish the viability of neurofeedback applications in prolonged weightless conditions.

Importantly, our paper could show that EEG, EMG, and eye tracking technologies could not only be used to estimate astronauts' cognitive and physical states but could also serve as a novel input modality for various technologies such as robotic systems and cabin equipment. In the nearer future, a hybrid sEMG- eye tracking-based system could be developed. Here, sEMG could be used to detect movement intentions during e.g. teleoperation, while eye tracking information could be used to adjust the angle of the camera or to change settings e.g. using a VR/AR head-mounted display. This strategy could be combined with (haptic) shared control schemes, in which the user and computer system collaborate to achieve a particular objective.

Overall, findings regarding the use of EEG-based BCIs indicate that while SSVEP-based BCIs already work relatively reliably and require less training time, the use of MI-based commands to control external technical devices appears to be most promising in the distant future, as it does not rely on external, potentially uncomfortable stimuli, but rather on mental commands from the user. Moreover, both technologies have been tested in weightless conditions [39], [49]. However, more research is required to allow, in particular, for higher accuracy and speed to improve the reliability and safety of such systems. Additional research should be conducted to expand the number of commands that can be communicated in real-world scenarios. Also in this regard, the use of ear-EEG technology has the potential to significantly improve usability during long-distance flights.

However, even though the proposed systems promise improved safety and performance during future human space flight missions, they only achieve real-world applicability once they have demonstrated superior usability and performance in direct comparison to conventional methods of operation (e.g., joystick, touch display). Notably, ethical concerns should be further evaluated, as the extended recordings of camera or brain activity may be perceived as a violation of personal privacy.

In the long run, combining technologies, such as EEG-based MI, eye-tracking, and sEMG, could be one potential solution to enhance the performance of these systems. This strategy could be furthermore combined with (haptic) shared control schemes. In this particular respect, [90] used MI (to modulate the movement speed of a robotic end effector) in conjunction with eye-tracking (to determine the movement direction) in an obstacle avoidance paradigm, aided by a computer that estimated the correct trajectory using computer vision techniques. They concluded that this combinational approach allows for high controllability and continuous, collision-free movement of the robotic system by the user while providing assistance in challenging situations (e.g. when obstacles are present). Similarly, other promising ideas, such as the smart astronaut glove developed by [91], could be enhanced by the proposed technologies. As of now, the glove integrates gesture-based commands with head- and eye-tracking technology, as well as an AR head-up display implemented in an astronaut suit, in order to intuitively communicate commands to a drone. In addition, the supplementation with other promising advanced technologies, such as voice control or fNIRS should be further evaluated to improve general system performance and usability.

**Acknowledgments**
We would like to thank all members of Spaceship EAC and the EAC XR – Lab for their comments and insightful discussions.

**References**
[1] A. Rafiq, R. Hummel, V. Lavrentyev, W. Derry, D. Williams, and R. C. Merrell, "Microgravity effects on fine motor skills: tying surgical knots

during parabolic flight," *Aviat. Space Environ. Med.*, vol. 77, no. 8, pp. 852–856, 2006.

[2] L. K. Barger *et al.*, "Prevalence of sleep deficiency and use of hypnotic drugs in astronauts before, during, and after spaceflight: an observational study," *Lancet Neurol.*, vol. 13, no. 9, pp. 904–912, 2014.

[3] C. R. Doarn, J. D. Polk, and M. Shepanek, "Health challenges including behavioral problems in long-duration spaceflight," *Neurol. India*, vol. 67, no. 8, p. 190, Jan. 2019, doi: 10.4103/0028-3886.259116.

[4] L. A. Palinkas, "Psychosocial issues in long-term space flight: overview," *Gravitational Space Res.*, vol. 14, no. 2, 2007.

[5] G. Petit *et al.*, "Local sleep-like events during wakefulness and their relationship to decreased alertness in astronauts on ISS," *Npj Microgravity*, vol. 5, no. 1, pp. 1–9, 2019.

[6] C. D. Wickens, "Mental Workload: Assessment, Prediction and Consequences," in *Human Mental Workload: Models and Applications*, Cham, 2017, pp. 18–29. doi: 10.1007/978-3-319-61061-0_2.

[7] M. R. Endsley and E. O. Kiris, "The out-of-the-loop performance problem and level of control in automation," *Hum. Factors*, vol. 37, no. 2, pp. 381–394, 1995.

[8] K. Holden, B. Munson, and J. Stephenson, "Monitoring Human Performance on Future Deep Space Missions," in *International Conference on Applied Human Factors and Ergonomics*, 2021, pp. 46–51.

[9] G. G. De la Torre, "Cognitive neuroscience in space," *Life*, vol. 4, no. 3, pp. 281–294, 2014.

[10] F. Carpi, D. De Rossi, and C. Menon, "Non invasive brain-machine interfaces," *ESA Ariadna Study*, vol. 5, p. 6402, 2006.

[11] M. Clerc, L. Bougrain, and F. Lotte, *Brain-Computer Interfaces 1*. Wiley-ISTE, 2016.

[12] T. Hummadi and I. Chatterjee, "An era of brain-computer interface: BCI migration into space," *Neurosci. Res. Notes*, vol. 3, no. 5, Art. no. 5, Jun. 2021.

[13] G. Clément and J. T. Ngo-Anh, "Space physiology II: adaptation of the central nervous system to space flight—past, current, and future studies," *Eur. J. Appl. Physiol.*, vol. 113, no. 7, pp. 1655–1672, Jul. 2013, doi: 10.1007/s00421-012-2509-3.

[14] Person and S. Scheer, "Israeli startup to test brain-activity gear on space mission to ISS," *Reuters*. Thomson Reuters, Mar. 2022. https://www.reuters.com/technology/israeli-startup-test-brain-activity-gear-space-mission-iss-2022-03-28/, (accessed 30.08.22).

[15] H. Cai, Z. Qu, Z. Li, Y. Zhang, X. Hu, and B. Hu, "Feature-level fusion approaches based on multimodal EEG data for depression recognition," *Inf. Fusion*, vol. 59, pp. 127–138, 2020.

[16] A. Dzedzickis, A. Kaklauskas, and V. Bucinskas, "Human Emotion Recognition: Review of Sensors and Methods," *Sensors*, vol. 20, no. 3, Art. no. 3, Jan. 2020, doi: 10.3390/s20030592.

[17] X. Hou, Y. Liu, O. Sourina, Y. R. E. Tan, L. Wang, and W. Mueller-Wittig, "EEG based stress monitoring," in *2015 IEEE International Conference on Systems, Man, and Cybernetics*, 2015, pp. 3110–3115.

[18] V. B. Pavlenko, S. V. Chernyi, and D. G. Goubkina, "EEG correlates of anxiety and emotional stability in adult healthy subjects," *Neurophysiology*, vol. 41, no. 5, pp. 337–345, 2009.

[19] D. C. Hammond, "Neurofeedback treatment of depression and anxiety," *J. Adult Dev.*, vol. 12, no. 2, pp. 131–137, 2005.

[20] M. Thompson and L. Thompson, "Neurofeedback for stress management," *Princ. Pract. Stress Manag.*, vol. 3, pp. 249–287, 2007.

[21] M. Basner *et al.*, "Mars 520-d mission simulation reveals protracted crew hypokinesis and alterations of sleep duration and timing," *Proc. Natl. Acad. Sci.*, vol. 110, no. 7, pp. 2635–2640, 2013.

[22] S. Pallesen and B. Bjorvatn, "3.2 WORKLOAD AND FATIGUE," *Space Saf. Hum. Perform.*, p. 71, 2017.

[23] H. Van Dongen, G. Maislin, J. M. Mullington, and D. F. Dinges, "The cumulative cost of additional wakefulness: dose-response effects on neurobehavioral functions and sleep physiology from chronic sleep restriction and total sleep deprivation," *Sleep*, vol. 26, no. 2, pp. 117–126, 2003.

[24] J. Bruun, "Smart earbud will measure how astronauts sleep," Jun. 2022. https://ingenioer.au.dk/en/current/news/view/artikel/danskdesignet-special-oereprop-skal-maale-astronauters-soevn, (accessed 30.08.22).

[25] A. Cortoos, E. De Valck, M. Arns, M. H. M. Breteler, and R. Cluydts, "An Exploratory Study on the Effects of Tele-neurofeedback and Tele-biofeedback on Objective and Subjective Sleep in Patients with Primary Insomnia," *Appl. Psychophysiol. Biofeedback*, vol. 35, no. 2, pp. 125–134, Jun. 2010, doi: 10.1007/s10484-009-9116-z.

[26] S. Sharma, S. K. Khare, V. Bajaj, and I. A. Ansari, "Improving the separability of drowsiness and alert EEG signals using analytic form of

wavelet transform," *Appl. Acoust.*, vol. 181, p. 108164, Oct. 2021, doi: 10.1016/j.apacoust.2021.108164.

[27] M. Hajinoroozi, Z. Mao, T.-P. Jung, C.-T. Lin, and Y. Huang, "EEG-based prediction of driver's cognitive performance by deep convolutional neural network," *Signal Process. Image Commun.*, vol. 47, pp. 549–555, Sep. 2016, doi: 10.1016/j.image.2016.05.018.

[28] Y. Zhou, S. Huang, Z. Xu, P. Wang, X. Wu, and D. Zhang, "Cognitive workload recognition using EEG signals and machine learning: A review," *IEEE Trans. Cogn. Dev. Syst.*, 2021.

[29] F. Atilla and M. Alimardani, "EEG-based Classification of Drivers Attention using Convolutional Neural Network," in *2021 IEEE 2nd International Conference on Human-Machine Systems (ICHMS)*, 2021, pp. 1–4.

[30] E. Dessy, M. Van Puyvelde, O. Mairesse, X. Neyt, and N. Pattyn, "Cognitive Performance Enhancement: Do Biofeedback and Neurofeedback Work?," *J. Cogn. Enhanc.*, vol. 2, no. 1, pp. 12–42, Mar. 2018, doi: 10.1007/s41465-017-0039-y.

[31] D. B. Kaber and M. R. Endsley, "The effects of level of automation and adaptive automation on human performance, situation awareness and workload in a dynamic control task," *Theor. Issues Ergon. Sci.*, vol. 5, no. 2, pp. 113–153, 2004.

[32] P. Aricò *et al.*, "Adaptive automation triggered by EEG-based mental workload index: a passive brain-computer interface application in realistic air traffic control environment," *Front. Hum. Neurosci.*, vol. 10, p. 539, 2016.

[33] S. T. Iqbal, X. S. Zheng, and B. P. Bailey, "Task-evoked pupillary response to mental workload in human-computer interaction," in *CHI '04 Extended Abstracts on Human Factors in Computing Systems*, New York, NY, USA, Apr. 2004, pp. 1477–1480. doi: 10.1145/985921.986094.

[34] D. Freer, Y. Guo, F. Deligianni, and G.-Z. Yang, "On-Orbit Operations Simulator for Workload Measurement during Telerobotic Training." arXiv, Jun. 22, 2020. doi: 10.48550/arXiv.2002.10594.

[35] C. Roos, J. Van der Pal, and G. Sewnath, "Maintaining optimal training value using Brain Computer Interfacing," 2017.

[36] X. Hou *et al.*, "EEG-based human factors evaluation of conflict resolution aid and tactile user interface in future air traffic control systems," in *Advances in Human Aspects of Transportation*, Springer, 2017, pp. 885–897.

[37] C. De Negueruela, M. Broschart, C. Menon, and J. del R. Millán, "Brain–computer interfaces for space applications," *Pers. Ubiquitous Comput.*, vol. 15, no. 5, pp. 527–537, 2011.

[38] X. Chen, B. Zhao, and X. Gao, "Noninvasive brain-computer interface based high-level control of a robotic arm for pick and place tasks," in *2018 14th International Conference on Natural Computation, Fuzzy Systems and Knowledge Discovery (ICNC-FSKD)*, 2018, pp. 1193–1197.

[39] S. Chen *et al.*, "An Experimental Study on Usability of Brain-Computer Interaction Technology in Human Spaceflight," in *Augmented Cognition. Enhancing Cognition and Behavior in Complex Human Environments*, Cham, 2017, pp. 301–312. doi: 10.1007/978-3-319-58625-0_22.

[40] L. Chu, J. Fernández-Vargas, K. Kita, and W. Yu, "Influence of stimulus color on steady state visual evoked potentials," in *International Conference on Intelligent Autonomous Systems*, 2016, pp. 499–509.

[41] K. El Batran and M. D. Dunlop, "Enhancing KLM (keystroke-level model) to fit touch screen mobile devices," in *Proceedings of the 16th international conference on Human-computer interaction with mobile devices & services*, New York, NY, USA, Sep. 2014, pp. 283–286. doi: 10.1145/2628363.2628385.

[42] Z. İşcan and V. V. Nikulin, "Steady state visual evoked potential (SSVEP) based brain-computer interface (BCI) performance under different perturbations," *PloS One*, vol. 13, no. 1, p. e0191673, 2018.

[43] S. Hayati and S. T. Venkataraman, "Design and implementation of a robot control system with traded and shared control capability," in *1989 IEEE International Conference on Robotics and Automation*, 1989, pp. 1310–1311.

[44] H.-T. Hsu, K.-K. Shyu, C.-C. Hsu, L.-H. Lee, and P.-L. Lee, "Phase-Approaching Stimulation Sequence for SSVEP-Based BCI: A Practical Use in VR/AR HMD," *IEEE Trans. Neural Syst. Rehabil. Eng.*, 2021.

[45] "Play, explore, develop with the power of your mind," *NextMind*. https://www.next-mind.com/, (accessed 30.08.22).

[46] Y. Yu *et al.*, "Toward brain-actuated car applications: Self-paced control with a motor imagery-based brain-computer interface," *Comput. Biol. Med.*, vol. 77, pp. 148–155, Oct. 2016, doi: 10.1016/j.compbiomed.2016.08.010.

[47] P. Arpaia, A. Esposito, A. Natalizio, and M. Parvis, "How to successfully classify EEG in motor imagery BCI: a metrological analysis of the state of the art," *J. Neural Eng.*, vol. 19, no. 3,

p. 031002, Jun. 2022, doi: 10.1088/1741-2552/ac74e0.

[48] S. Brandl, J. Höhne, K.-R. Müller, and W. Samek, "Bringing BCI into everyday life: Motor imagery in a pseudo realistic environment," in *2015 7th International IEEE/EMBS Conference on Neural Engineering (NER)*, 2015, pp. 224–227.

[49] J. del R. Millàn, P. W. Ferrez, and T. Seidl, "Chapter 14 Validation of Brain–Machine Interfaces During Parabolic Flight," in *International Review of Neurobiology*, vol. 86, Academic Press, 2009, pp. 189–197. doi: 10.1016/S0074-7742(09)86014-5.

[50] A. Singh, A. A. Hussain, S. Lal, and H. W. Guesgen, "A Comprehensive Review on Critical Issues and Possible Solutions of Motor Imagery Based Electroencephalography Brain-Computer Interface," *Sensors*, vol. 21, no. 6, Art. no. 6, Jan. 2021, doi: 10.3390/s21062173.

[51] E. B. J. Coffey, A.-M. Brouwer, E. S. Wilschut, and J. B. F. van Erp, "Brain–machine interfaces in space: Using spontaneous rather than intentionally generated brain signals," *Acta Astronaut.*, vol. 67, no. 1, pp. 1–11, Jul. 2010, doi: 10.1016/j.actaastro.2009.12.016.

[52] G. Gillini, P. D. Lillo, and F. Arrichiello, "An Assistive Shared Control Architecture for a Robotic Arm Using EEG-Based BCI with Motor Imagery," in *2021 IEEE/RSJ International Conference on Intelligent Robots and Systems (IROS)*, Sep. 2021, pp. 4132–4137. doi: 10.1109/IROS51168.2021.9636261.

[53] P. Majaranta and A. Bulling, "Eye Tracking and Eye-Based Human–Computer Interaction," in *Advances in Physiological Computing*, S. H. Fairclough and K. Gilleade, Eds. London: Springer, 2014, pp. 39–65. doi: 10.1007/978-1-4471-6392-3_3.

[54] J. Z. Lim, J. Mountstephens, and J. Teo, "Emotion Recognition Using Eye-Tracking: Taxonomy, Review and Current Challenges," *Sensors*, vol. 20, no. 8, Art. no. 8, Jan. 2020, doi: 10.3390/s20082384.

[55] D. Martinez-Marquez, S. Pingali, K. Panuwatwanich, R. A. Stewart, and S. Mohamed, "Application of eye tracking technology in aviation, maritime, and construction industries: a systematic review," *Sensors*, vol. 21, no. 13, p. 4289, 2021.

[56] F. Di Nocera, M. Camilli, and M. Terenzi, "A random glance at the flight deck: Pilots' scanning strategies and the real-time assessment of mental workload," *J. Cogn. Eng. Decis. Mak.*, vol. 1, no. 3, pp. 271–285, 2007.

[57] L. S. Stone, T. L. Tyson, P. F. Cravalho, N. H. Feick, and E. E. Flynn-Evans, "Distinct pattern of oculomotor impairment associated with acute sleep loss and circadian misalignment," *J. Physiol.*, vol. 597, no. 17, pp. 4643–4660, 2019.

[58] T. Bafna and J. P. Hansen, "Mental fatigue measurement using eye metrics: A systematic literature review," *Psychophysiology*, vol. 58, no. 6, p. e13828, 2021.

[59] F. Li, C.-H. Chen, G. Xu, and L.-P. Khoo, "Hierarchical Eye-Tracking Data Analytics for Human Fatigue Detection at a Traffic Control Center," *IEEE Trans. Hum.-Mach. Syst.*, vol. 50, no. 5, pp. 465–474, Oct. 2020, doi: 10.1109/THMS.2020.3016088.

[60] M. Q. Khan and S. Lee, "Gaze and Eye Tracking: Techniques and Applications in ADAS," *Sensors*, vol. 19, no. 24, Art. no. 24, Jan. 2019, doi: 10.3390/s19245540.

[61] J. G. Tichon, T. Mavin, G. Wallis, T. A. W. Visser, and S. Riek, "Using pupillometry and electromyography to track positive and negative affect during flight simulation," *Aviat. Psychol. Appl. Hum. Factors*, vol. 4, no. 1, pp. 23–32, 2014, doi: 10.1027/2192-0923/a000052.

[62] J. Allsop and R. Gray, "Flying under pressure: Effects of anxiety on attention and gaze behavior in aviation," *J. Appl. Res. Mem. Cogn.*, vol. 3, no. 2, pp. 63–71, 2014.

[63] S. Peißl, C. D. Wickens, and R. Baruah, "Eye-tracking measures in aviation: A selective literature review," *Int. J. Aerosp. Psychol.*, vol. 28, no. 3–4, pp. 98–112, 2018.

[64] K. Kilingaru, J. W. Tweedale, S. Thatcher, and L. C. Jain, "Monitoring pilot 'situation awareness,'" *J. Intell. Fuzzy Syst.*, vol. 24, no. 3, pp. 457–466, 2013.

[65] S. Piechowski et al., "Visual Attention Relates to Operator Performance in Spacecraft Docking Training," *Aerosp. Med. Hum. Perform.*, vol. 93, no. 6, pp. 480–486, Jun. 2022, doi: 10.3357/AMHP.6005.2022.

[66] V. A. Huemer et al., "Characterizing scan patterns in a spacecraft cockpit simulator: expert vs. novice performance," in *Proc. of the Human Factors and Ergonomics Society Annual Meeting*, 2005, vol. 49, no. 1, pp. 83–87.

[67] Y. Guo, D. Freer, F. Deligianni, and G.-Z. Yang, "Eye-Tracking for Performance Evaluation and Workload Estimation in Space Telerobotic Training," *IEEE Trans. Hum.-Mach. Syst.*, vol. 52, no. 1, pp. 1–11, Feb. 2022, doi: 10.1109/THMS.2021.3107519.

[68] C. Diaz-Piedra, H. Rieiro, A. Cherino, L. J. Fuentes, A. Catena, and L. L. Di Stasi, "The effects of flight complexity on gaze entropy: An experimental study with fighter pilots," *Appl. Ergon.*, vol. 77, pp. 92–99, May 2019, doi:

10.1016/j.apergo.2019.01.012.
[69] D. St-Onge *et al.*, "Planetary exploration with robot teams," *IEEE Robot. Autom. Mag.*, 2019.
[70] T. D. Greef, H. Lafeber, H. V. Oostendorp, and J. Lindenberg, "Eye Movement as Indicators of Mental Workload to Trigger Adaptive Automation," *Found. Augment. Cogn. Neuroergonomics Oper. Neurosci. Lect. Notes Comput. Sci.*, pp. 219–228, 2009, doi: 10.1007/978-3-642-02812-0_26.
[71] Y. Tian, S. Chen, C. Wang, Q. Yan, and Z. Wang, "Investigations on eye movement activities in the manually controlled rendezvous and docking of space vehicles," *Adv. Phys. Ergon. Saf.*, pp. 182–189, 2012.
[72] R. Alonso, M. Causse, F. Vachon, R. Parise, F. Dehais, and P. Terrier, "Evaluation of head-free eye tracking as an input device for air traffic control," *Ergonomics*, vol. 56, no. 2, pp. 246–255, Feb. 2013, doi: 10.1080/00140139.2012.744473.
[73] J. Chen, Z. Ji, H. Niu, R. Setchi, and C. Yang, "An auto-correction teleoperation method for a mobile manipulator using gaze tracking and hand motion detection," in *Annual Conference Towards Autonomous Robotic Systems*, 2019, pp. 422–433.
[74] L. Bi and C. Guan, "A review on EMG-based motor intention prediction of continuous human upper limb motion for human-robot collaboration," *Biomed. Signal Process. Control*, vol. 51, pp. 113–127, 2019.
[75] M. S. Rashid, Z. Zaman, H. Mahmud, and M. K. Hasan, "Emotion Recognition with Forearm-based Electromyography." arXiv, Nov. 13, 2019. doi: 10.48550/arXiv.1911.05305.
[76] M. Zhuang *et al.*, "Highly robust and wearable facial expression recognition via deep-learning-assisted, soft epidermal electronics," *Research*, vol. 2021, 2021.
[77] S. Pourmohammadi and A. Maleki, "Stress detection using ECG and EMG signals: A comprehensive study," *Comput. Methods Programs Biomed.*, vol. 193, p. 105482, Sep. 2020, doi: 10.1016/j.cmpb.2020.105482.
[78] A. Canter, C. Y. Kondo, and J. R. Knott, "A Comparison of EMG Feedback and Progressive Muscle Relaxation Training in Anxiety Neurosis," *Br. J. Psychiatry*, vol. 127, no. 5, pp. 470–477, Nov. 1975, doi: 10.1192/bjp.127.5.470.
[79] F. N. Biondi, A. Cacanindin, C. Douglas, and J. Cort, "Overloaded and at Work: Investigating the Effect of Cognitive Workload on Assembly Task Performance," *Hum. Factors*, vol. 63, no. 5, pp. 813–820, Aug. 2021, doi: 10.1177/0018720820929928.
[80] T. P. Moore, "US space flight experience. Physical exertion and metabolic demand of extravehicular activity: Past, present, and future," 1989.
[81] R. Di Giminiani *et al.*, "A wearable integrated textile EMG and muscle oximetry system for monitoring exercise-induced effects: a feasibility study," in *2018 IEEE International Symposium on Medical Measurements and Applications (MeMeA)*, Jun. 2018, pp. 1–5. doi: 10.1109/MeMeA.2018.8438785.
[82] C. Di Natali *et al.*, "Quasi-Passive Resistive Exosuit for Space Activities: Proof of Concept," *Appl. Sci.*, vol. 11, no. 8, Art. no. 8, Jan. 2021, doi: 10.3390/app11083576.
[83] A. Hagengruber, U. Leipscher, B. M. Eskofier, and J. Vogel, "Electromyography for Teleoperated Tasks in Weightlessness," *IEEE Trans. Hum.-Mach. Syst.*, vol. 51, no. 2, pp. 130–140, Apr. 2021, doi: 10.1109/THMS.2020.3047975.
[84] H. F. Hassan, S. J. Abou-Loukh, and I. K. Ibraheem, "Teleoperated robotic arm movement using electromyography signal with wearable Myo armband," *J. King Saud Univ.-Eng. Sci.*, vol. 32, no. 6, pp. 378–387, 2020.
[85] L. B. Jaloza, "Inside Facebook Reality Labs: Wrist-based interaction for the next computing platform," *Tech at Meta*, Mar. 18, 2021. https://tech.fb.com/ar-vr/2021/03/inside-facebook-reality-labs-wrist-based-interaction-for-the-next-computing-platform/, (accessed 30.08.22).
[86] J. Rebelo, T. Sednaoui, E. B. den Exter, T. Krueger, and A. Schiele, "Bilateral Robot Teleoperation: A Wearable Arm Exoskeleton Featuring an Intuitive User Interface," *IEEE Robot. Autom. Mag.*, vol. 21, no. 4, pp. 62–69, Dec. 2014, doi: 10.1109/MRA.2014.2360308.
[87] J. C. Mankins, "Technology readiness levels," *White Pap. April*, vol. 6, no. 1995, p. 1995, 1995.
[88] D.-H. Jeong and J. Jeong, "In-Ear EEG Based Attention State Classification Using Echo State Network," *Brain Sci.*, vol. 10, no. 6, Art. no. 6, Jun. 2020, doi: 10.3390/brainsci10060321.
[89] R. van Lutterveld *et al.*, "Source-space EEG neurofeedback links subjective experience with brain activity during effortless awareness meditation," *NeuroImage*, vol. 151, pp. 117–127, May 2017, doi: 10.1016/j.neuroimage.2016.02.047.
[90] H. Zeng *et al.*, "Semi-Autonomous Robotic Arm Reaching With Hybrid Gaze–Brain Machine Interface," *Front. Neurorobotics*, vol. 13, 2020.
[91] P. Lee *et al.*, "Astronaut Smart Glove: A Human-Machine Interface For the Exploration of the Moon, Mars, and Beyond," 2020.